# Predictive AI for SME and Large Enterprise Financial Performance Management

Ricardo Cuervo (MBA)
*MSc. Artificial Intelligence, QMUL*
rcuervo@manacorgroup.com

*Abstract*—Financial performance management is at the core of business management and has historically relied on financial ratio analysis using Balance Sheet and Income Statement data to assess a company's performance vis-à-vis competitors. Little progress has been made in predicting how a company will perform or in assessing the risks (probabilities) of financial underperformance. In this study I introduce a new set of financial and macroeconomic ratios that supplement standard ratios of Balance Sheet and Income Statement. I also provide a set of supervised learning models (ML Regressors and Neural Networks) and Bayesian models to predict a company's performance. I conclude that the new proposed variables improve model accuracy when used in tandem with standard industry ratios. I also conclude that Feedforward Neural Networks (FNN) are simpler to implement and perform best across 6 predictive tasks (ROA, ROE, Net Margin, Op Margin, Cash Ratio and Op Cash Generation); although Bayesian Networks (BN) can outperform FNN under very specific conditions. BNs have the additional benefit of providing a probability density function in addition to the predicted (expected) value. The study findings have significant potential helping CFOs and CEOs assess risks of financial under-performance to steer companies in more profitable directions; supporting lenders in better assessing a company's condition and providing investors with tools to dissect public companies' financial statements more accurately.

## I. Introduction

Financial Ratio Analysis is "a quantitative method of gaining insight into a company's liquidity, operational efficiency, and profitability by studying its financial statements such as the balance sheet and income statement. Ratio analysis is a cornerstone of fundamental equity analysis" (Bloomenthal, 2023). The study of a company's performance through financial ratios, spanning decades, is largely characterized by two main factors:

(1) ) Financial Ratio Analysis significantly focuses on ratios derived from Balance Sheet and Income Statement. See **Table A1** for an extensive list of ratios normally used.

(2) Financial Ratio Analysis is often used comparatively to determine the overall strength of a company's financial health. As such, analyses are usually of one of three types:

- Financial ratio analysis across companies, to determine relative health vis-à-vis competitors.
- Financial ratio analysis against benchmarks, to set targets for improvement.
- Financial ratio analysis over time, to determine improving vs. deteriorating trends.

Four main observations arise:

1- Approach does not consider Statement of Cash Flows despite cash flows being a fundamental part of a company's financial health. Companies can be profitable but illiquid, jeopardizing their business prospects. Furthermore, better valuation approaches favour free cash flows over profits.

2- Approach does not account for varying macroeconomic conditions. Static financial ratio analysis based on historical benchmarks without properly accounting for a changing macroeconomic context is, consequently, severely limited.

3- Predictive modelling in financial ratio analysis has been significantly limited to time series analysis, forecasting future performance based on the trend of a ratio (e.g. autoregressive moving average models), with limited consideration of cross-dependencies between financial indicators.

4- A typical decision by a CEO simultaneously impacts multiple accounts in the company's financial statements. However, a financial ratio analysis that looks into each metric separately fails to account for such interdependencies. For example, a decision to discount inventories may increase sales in the short term, improve inventory turnover, affect current ratio and increase operating cashflows, while negatively impacting operating margin. Similarly, a decision to extend payment terms may also increase sales, increase accounts payable, positively impact operating margin while reducing operating cashflows. The standard approach of looking separately into liquidity, profitability, leverage and efficiency ratios relies heavily on the skills of the financial analyst to learn such interdependencies.

More advanced financial modelling methodologies incorporating Machine Learning (ML) and deep learning can be found mostly in commercial banking credit operations as well as in investment applications. However, bank models emphasize credit risk variables and their proprietary nature makes them inaccessible to a broader audience.

## II. Problem Statement

Can a company's financial performance be predicted for the next business cycle (quarter) based on its current state? If we accept that the financial statements (Balance Sheet, Income Statement, Statement of Cash Flows, and accompanying notes to financial statements) provide the required information that best illustrates the company's state of affairs, the task becomes to determine the appropriate financial metrics and to design, train and test the model(s) that best predict company's outcome.

Predicting *next quarter financial performance* based on current financial data implies that a company's performance has the *Markov property*, namely that the current state summarizes the entire past with all information that is relevant for decision making and by extension, for prediction of future state. This approach contrasts with time series forecasting where historical data is used for predictions. This approach also enables incorporating exogenous macroeconomic data: Afterall, the current state for business decision making is not limited to endogenous company data.

The problem of assessing and predicting financial performance is prevalent in the industry. The ability to make accurate predictions about a company's financial performance would increase market efficiency for investors, reduce company-specific idiosyncratic risk for lenders and provide CEOs and CFOs with tools to improve overall business outcomes and returns for stakeholders.

## III. CONTRIBUTION

This project makes the following contributions to the field of equity analysis and financial performance management:

First, I introduce nine new financial ratios that focus on Statement of Cash Flows (CF) and macroeconomic metrics, which supplement ratios typically used based on Balance Sheet (BS) and Income Statement (IS). I find that these new ratios significantly augment predictive capabilities, yielding improved performance of supervised learning models (ML Regressors and Feedforward Neural Networks -FNN-) by as much as 11% compared to similar models using standard Balance Sheet and Income Statement metrics only.

Second, I propose a methodology to develop, train and evaluate predictive financial performance models, by which quarter statement data (multiple independent variables) is tagged with next quarter performance metrics (dependent variables). Models trained with this data learn to make predictions based on the company's financial ratios at a point in time. This will allow a financial analyst to formulate predictions about a company's performance based on current financial statement and macroeconomic data (Markov property).

Third, I propose a set of ML Regression and FNN models to predict financial performance across 6 different target metrics (*ROA, ROE, Net Margin, EBIT, Cash Ratio and Operating Cash Generation*). I find that FNNs outperform ML regressors and ensemble models by as much as 26%.

Fourth, I propose a causal Bayesian Network (BN) model to predict financial performance with a latent unobservable *business performance* node intended to capture overall *quality of business execution*, conditioned by a set of macroeconomic variables. Since this node is unobservable, its probability distribution cannot be explicitly stated. Instead, five indicator nodes -instantiations of measurement idioms- are used to observe business performance. The final node of the model corresponds to the outcome variable, e.g. Predicted Net Margin. It is influenced by the macroeconomic variables and by the unobservable quality of business execution. I find that within a narrow range of expected business performance close to industry mean values, the Bayesian Network model can outperform FNN models. However, its accuracy is diluted across a wider range of companies placed farther away from industry averages. While the Node Probability Table -NPT- distributions and parent/child relationships can be further evaluated and potentially improved, the time-consuming nature of the BN model design and experimentation process limits its applicability vis-à-vis supervised learning approaches.

## IV. BACKGROUND / RELATED WORK

Generally speaking, AI applications in corporate finance are of two types: *numerical data* (structured, e.g., stock prices, financial statements, time-series data) and *text data* (unstructured, e.g., financial news, disclosures, notes to financial statements).

Specifically on numerical data, research on deep learning methods or corporate finance applications is largely focused in two areas: (1) commercial banking and lending (*credit risk and bankruptcy risk*), and (2) investments (*stock price*). Huang et al. (2020) provide a thorough review of research on deep learning in finance and banking, identifying and cataloguing 59 projects deemed of 'higher quality' conducted during the period 2014 – 2018 from an initial list of more than 150 cross-referenced articles. Specifically, they identified 7 finance and banking domains for the application of deep learning (**Fig. 1**). It should be noticed that business performance management as a DL domain is missing. Their work updates previous research by Cavalcante et al. (2016), who provided a comprehensive survey of ML methods applied to the financial context from 2009 to 2015.

Hosaka (2018) proposes an innovative approach applying convolutional neural networks on financial ratios data to predict bankruptcy risk. While the scope is limited to bankruptcy prediction, the approach covers a wide business management spectrum, capturing multitude of financial ratios from 133 statement accounts in BS and IS. The proposed approach contrasts *Altman Z-score* handcrafted feature selection, taking instead all possible ratios constructed with the BS and IS statements, and using a CNN to learn the data and determine features of higher relevance in predicting bankruptcy risk. However, the use of convolutions to extract features from localized regions in an input data array does not seem applicable to extract features from a set of numerical financial data (financial statements) where the order (location) of the data in the array is not relevant. FNN with hidden layers connected to all input variables are more appropriate. Indeed, the author concluded that "the proposed method is not suitable for the purpose of investigating the causes of bankruptcy".

Chakraborty (2019) proposes a reinforcement learning approach to capture financial market trading data, which incorporates a Markov decision process approach to trade profitably. His research recognizes the challenges of getting reliable, exhaustive, high-quality data to effectively enable deep learning to find optimal policies. Instead, it uses a set of *technical indicators*, which are mathematical and statistical data calculated from the time-series pricing data of the instrument. The use of such time-dependent data to model a Markov decision process appears to be contradictory in nature.

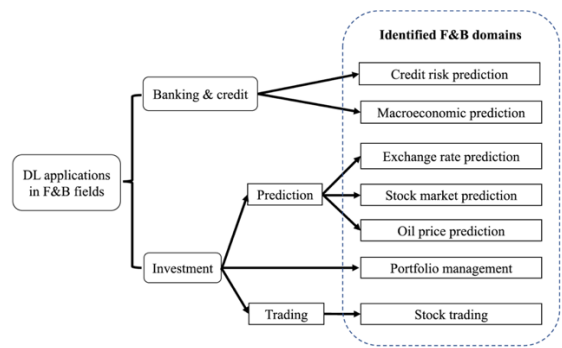

Fig. 1. Deep Learning Applications in Finance and Banking. Huang et al. (2020).

Ample research has been conducted on the use of machine learning (ML), reinforcement learning (RL) and deep learning (DL) models to optimize investment strategies using stock price as solely indicator of a company's financial performance (e.g., *DL*: Kim et al., 2023; *RL*: Liu et al., 2022; Liang et al., 2018; *ML*: Yang, Liu and Wu, 2018). The Efficient Market Hypothesis (Malkiel, 2003) establishes that stock prices fully reflect all known information and thus, markets do not allow investors to earn above-average returns without accepting above-average risks. However, approaches to obtain above-market returns based solely on stock price data fail to capture endogenous company data as they rely only on how markets perceive it. It can be argued that there is a lag between a company undertaking an action and the market knowing it.

Deep learning methods using text data have also gained traction in finance with applications appearing to be mostly focused on same financial domains. Several articles on investor behaviour research and trading systems based on investor sentiment have emerged in current literature. Mishev et al. (2020) present a comprehensive chronological study of NLP-based methods for sentiment analysis in finance, from word and sentence encoders to NLP transformers. They conclude that NLP transformers show superior performance. Yang et. al. (2018) propose a trading system based on investor sentiment, trained with news sentiment from Thomson Reuters news analytics database. Other relevant work includes Checkley, Añón Higón, and Alles (2017), Oliveira, Cortez, and Areal (2017), Day and Lee (2016), Feuerriegel and Prendinger (2018). Nevertheless, a similar observation can be made: By focusing on use of exogenous data (i.e., news sentiment), such approaches do not capture endogenous data that may provide rich insights from company's insiders and improve understanding of a company's financial performance.

## V. METHODOLOGY

### A. Datasets and Data Preprocessing

Public companies in USA are required to file financial statements with the Securities and Exchange Commission (SEC) every quarter (10Q filings) and annually (10K) among other financial disclosure requirements. This data is accessible through the SEC EDGAR system. A total of 461k filings were made by 13655 non-financial services companies during the period Jan 2010 – Apr 2023. Of these, 265k correspond to 10Q and 10K reports. See **Table A2** for a breakdown by industry.

While accounting principles and regulations pertaining preparation of financial statements are very well defined, data is not homogeneous across companies. Thus, as part of this research, a significant effort was undertaken to extract, cleanse and standardize financial statement data, and annotate it with the to-be-predicted target variables. Examples of SEC data cleansing and preparation challenges include:

- Tags are not consistently named or used. There are over 500k different tags in the SEC dataset.

- Polarity of certain numbers is not consistent. For example, a company may report an expense as a positive number to be subtracted from revenues. Another company may report the same expense as a negative number to be 'added' to the revenues to obtain gross profit.

- Some companies report cumulative quarterly data (e.g., Q3 values comprise 9 months of operations), while others may report strict 3-months data irrespective of quarter.

Data was extracted into standardized financial statement templates (**Fig. A1**). A clean data set of financial ratios was then calculated using the templatized financial statements data. A dataset with a total of 61K clean records was produced. Next step in data preparation consisted of statistical analysis to detect and eliminate outliers and 'noisy data'. Further data cleansing and preparation yielded a final dataset of 46,141 records with templatized financial statement data to be used for model training and evaluation.

### B. Financial Ratios and Metrics

A total of 43 variables (financial ratios from the templatized financial statements and macroeconomic indexes) are calculated for each observation in the 46k dataset:

- 24 ratios form BS and IS data, which are most commonly used. These belong to 4 groups in financial ratio analysis literature: liquidity, profitability, asset utilization and financial leverage. See **Table A1**.

- Additional 6 new ratios from the Statement of Cash Flows (CF), intended to assess the quality of the company's cashflows. These are: Net Cash Generation (NCG), Operating Cash Generation (OCG), Current Liabilities Cash Coverage (CLCC), Operating Cash to Sales Ratio (OCS), Quality of Payment Terms (QPT) and Quality of Op. to Fin. Funds Use Ratio (QOFF). See **Table A3**.

- Additional 3 new macroeconomic-related financial statement (MRFS) ratios, which relate financial statement data to macroeconomic indexes and are intended to assess company's operations in changing economic environment. These are Liabilities to Yield Curve Alignment (LYCA), Inflation-Adjusted Inventory Carry-On Cost (IAICOC) and ROA to Bond Rate Ratio (ROA2bond). See **Table A3**.

- 10 macroeconomic indexes (exogenous variables), to capture the economic environment under which a company performs. See **Fig. A2.**

The subset of 6 CF financial ratios plus the 3 MRFS ratios (**Table A3**) constitute the set of 9 new ratios proposed under current research.

***Dependent Variables***: Each observation corresponds to the financial statements of a public company at different points in time ($t$). Data is augmented (tagged) with 6 to-be-predicted target variables, which correspond to the company's performance in the following quarter ($t+1$). These are ROA, ROE, Net Margin, Op. Margin, Cash Ratio and OCG.

The data shows low correlations across ratios, with few exceptions. **Fig. A3** shows the correlation matrix of all financial ratios used. Low correlation values make the modelling more feature-rich and thus, feature selection under different scenarios is expected to yield variances in model performance.

### C. ML Regression for Predictive Financial Modelling

3 types of models were considered to predict financial performance: ML Regressors, Neural Networks and Bayesian Networks. As it pertains to ML regressors, 6 model classes were evaluated: Linear Regression, Lasso, Elastic Net, KNN, Decision Tree Regressors (CART) and SVR.

<u>*Predictive Tasks*</u>: 6 different tasks were implemented to predict the following target variables: ROA, ROE, Net Margin, Operating Margin, Cash Ratio and OCG Ratio

(Operating Cash Generation). Selection of these variables as company's performance indicators was based on need to cover profitability, liquidity, and asset utilization (turnover) ratios. These are best metrics to assess overall performance at-a-glance and are widely used in the equity analysis industry.

*Scenarios*: Each model was trained using four different sets of variables to assess which ones have more predictive power and to determine impact from proposed new 9 cash flow and MRFS financial ratios:

- **base**: Base scenario refers to use of standard BS and IS ratios only. The purpose of this scenario is to baseline model performance using the standard metrics, which narrows analysis to balance sheet and income statement. 24 ratios were considered.

- **all variables**: Uses the entire set of 43 metrics including all ratios from BS and IS (24), CF (6), MRFS (3) plus macroeconomic indexes (10). The purpose of this scenario is to assess impact on model performance from the new metrics being proposed, and the exogenous macroeconomic variables, which are not controlled by a company's management.

- **fin-st vars**: Similar to previous scenario but excluding exogenous macroeconomic variables. This scenario allows us to isolate impact on financial performance from the new set of proposed metrics. A total of 33 BS, IS, CF, MRFS ratios is used for model training. Although no macroeconomic indexes are included, macroeconomic-related financial statement ratios (MRFS) do capture macroeconomic data.

- **new vars**: In this scenario, models are trained using only the new 9 CF and MRFS ratios proposed. This is to determine if the new metrics are enough to predict financial performance or if they work better supplementing standard metrics.

### D. Neural Networks for Predictive Financial Modelling

Four Feedforward Neural Networks were implemented for the same 6 predictive tasks as in ML Regression (ROA, ROE, Net Margin, Operating Margin, Cash Ratio and OCG Ratio), and trained with the same 4 sets of variables (scenarios).

- **Base Model**: A single hidden layer of size equal to the number of variables for each scenario (24, 43, 33, 9), followed by the output layer of size 1 corresponding to the target variable of each predictive task.

- **Deep Model**: Two hidden layers, the first one of size equal to number of variables for each scenario, followed by a layer of size 20.

- **Wide Model**: A single hidden layer of size 100.

- **Deep and Wide**: Two hidden layers (100 and 20)

All FNNs were implemented using sigmoid activation, Adam optimizer and MSE loss function, to provide for a fair comparison across network architectures and scenarios. Model hyperparameter optimization is not examined under this research. Models are trained and evaluated using 10-fold evaluation. Observed results do not suggest a tendency to overfit the data. Thus, neither convolutional neural networks nor other regularization techniques appear to be required as an alternative to fully connected networks.

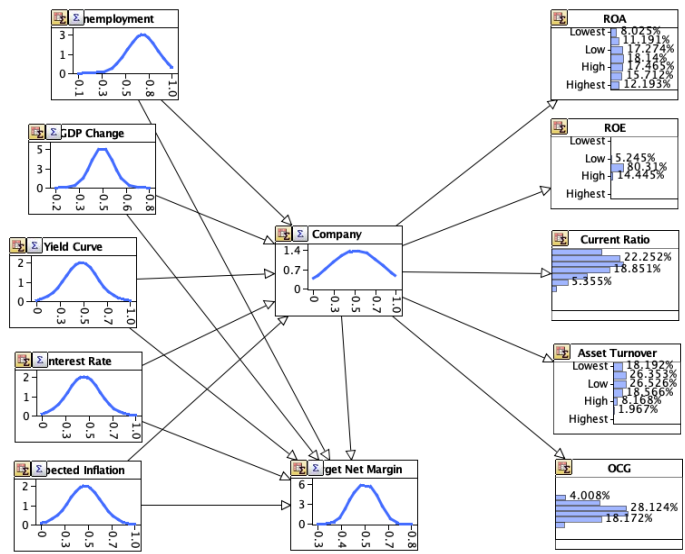

Fig. 2. Causal BN to determine Net Margin expected value and node PDF.

### E. Bayesian Network Probabilistic Financial Performance Modelling

A Bayesian causal network model predicting Net Margin is implemented in Agena.ai platform (**Fig. 2**). The business outcome node represents the target variable *Net Margin*. It depends on the overall company performance and is conditioned by the macroeconomic environment. Thus, the outcome node to be predicted is implemented as child to a set of parent macroeconomic nodes and to a node that encapsulates the quality of business execution (*Company Performance*). This node is latent and unobservable, just as it is in real life,. Consequently, leveraging the concept of measurement idioms, a set of indicator nodes are implemented to capture observable business performance metrics: ROA, ROE, Current Ratio, Asset Turnover and OCG. These are 5 variables selected from the set of 43 variables used in the training of the ML Regression and FNN models. This selection is somehow arbitrary, although intended to cover representative metrics of profitability, liquidity and asset utilization (turnover). Limiting the number of indicator nodes reduces complexity of the model and ensures that NPTs (Node Probability Tables) can be consistently calculated.

The parent macroeconomic nodes are Unemployment Rate, GDP change, Yield Curve, 10-YR Interest Rate and Expected Inflation. These nodes provide a good picture of the economic environment under which a company must perform. These 5 metrics evidence low correlations (**Fig. 3**), supporting overall assumption of node independence in the causal BN. The implementation of the BN follows a different approach than ML Regression or FNN models. While the 46k dataset is used to learn the parameters of the supervised learning models, the data statistics are used to determine the NPTs in the BN.

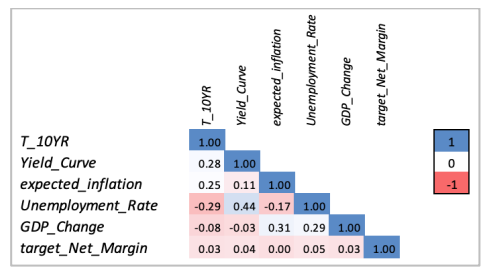

Fig. 3. Correlation Coefficients between macroeconomic indexes and target variable used in Bayesian Network model.

The business outcome node (*target Net Margin*) and the macroeconomic nodes are implemented as continuous T-Normal probability distributions. The business performance indicator nodes are implemented as ranked nodes with various number of states (from 7 to 11) and NPTs that closely follow the shape of the distributions from the 46K dataset.

All nodes are implemented in the range 0:1. Thus, the actual values are min-max normalized before calculating the underlying T-Normal parameters (mean and std-dev). Two exceptions apply:

- *Unemployment*: rather than using unemployment rates, the model takes employment rates. This is to align it with the concept that values closer to 1 reflect better macroeconomic condition.
- *OCG*: Operating Cash Generation is one of the new variables proposed under current research. The values in the 46K training dataset are in the range -41 to +46. Large absolute values are characteristic of business transient states and not sustainable over time. Thus, the shape of the distribution histogram is very narrow with very long tails (i.e. very low variance). Use of ranked nodes would place most of the companies in the middle bucket, not allowing for proper differentiation across companies. A sigmoid transform is used to 'stretch' the values around the mean, to enable more proper differentiation across companies (**Fig. A4**).

## VI. RESULTS

### A. ML Regression for Predictive Financial Modelling

Models were evaluated using MSE as loss metric. Model performance during training is evaluated using 10-fold cross-validation, which generally results in less biased estimates of model performance, especially considering the presence of some companies with extreme values for some of the financial ratios. **Table 1** presents 10-fold average MSE across all six predictive tasks (*target variables*), for all 4 scenarios (*feature sets*), for the six regression models considered.

TABLE 1. ML REGRESSION – MODEL PERFORMANCE RESULTS

| Predictive Task | Feature Set (Scenario) | Linreg | Lasso | ElasNet | KNN | CART | SVR |
|---|---|---|---|---|---|---|---|
| ROA | all variables | 0.0013 | 0.0021 | 0.0021 | 0.0019 | 0.0027 | 0.0020 |
| ROA | base | 0.0013 | 0.0021 | 0.0021 | 0.0021 | 0.0027 | 0.0026 |
| ROA | fin-st vars | 0.0013 | 0.0021 | 0.0021 | 0.0018 | 0.0028 | 0.0020 |
| ROA | new vars | 0.0014 | 0.0021 | 0.0021 | 0.0015 | 0.0028 | 0.0018 |
| ROE | all variables | 0.0174 | 0.0258 | 0.0258 | 0.0240 | 0.0345 | 0.0209 |
| ROE | base | 0.0175 | 0.0258 | 0.0258 | 0.0263 | 0.0345 | 0.0227 |
| ROE | fin-st vars | 0.0174 | 0.0258 | 0.0258 | 0.0237 | 0.0340 | 0.0207 |
| ROE | new vars | 0.0204 | 0.0258 | 0.0258 | 0.0231 | 0.0413 | 0.0200 |
| Net Margin | all variables | 0.0568 | 0.1014 | 0.0934 | 0.0936 | 0.1139 | 0.0709 |
| Net Margin | base | 0.0580 | 0.1014 | 0.1004 | 0.1027 | 0.1185 | 0.0774 |
| Net Margin | fin-st vars | 0.0569 | 0.1014 | 0.0935 | 0.0891 | 0.1147 | 0.0683 |
| Net Margin | new vars | 0.0677 | 0.1027 | 0.0940 | 0.0703 | 0.1248 | 0.0615 |
| Op. Margin | all variables | 0.0482 | 0.0975 | 0.0882 | 0.0861 | 0.0949 | 0.0641 |
| Op. Margin | base | 0.0494 | 0.0976 | 0.0965 | 0.0955 | 0.0982 | 0.0700 |
| Op. Margin | fin-st vars | 0.0483 | 0.0976 | 0.0885 | 0.0814 | 0.0949 | 0.0611 |
| Op. Margin | new vars | 0.0612 | 0.0990 | 0.0891 | 0.0625 | 0.1056 | 0.0544 |
| Cash Ratio | all variables | 0.2791 | 1.0082 | 0.6237 | 0.8093 | 0.5831 | 0.4647 |
| Cash Ratio | base | 0.2808 | 1.0082 | 0.6237 | 0.7500 | 0.5904 | 0.4120 |
| Cash Ratio | fin-st vars | 0.2802 | 1.0082 | 0.6237 | 0.7474 | 0.5963 | 0.4124 |
| Cash Ratio | new vars | 1.0224 | 1.1131 | 1.1131 | 0.7180 | 0.6246 | 0.8338 |
| OCG | all variables | 7.2902 | 7.6285 | 7.5269 | 7.9628 | 13.5663 | 7.8628 |
| OCG | base | 8.2151 | 8.5832 | 8.5795 | 8.7221 | 12.3635 | 8.5454 |
| OCG | fin-st vars | 7.2926 | 7.6288 | 7.5272 | 7.5410 | 13.5121 | 7.5368 |
| OCG | new vars | 7.3676 | 7.6609 | 7.5563 | 7.6785 | 13.4202 | 6.9191 |

TABLE 2. BEST ML REGRESSION MODEL

| Predictive Task | Best Performing Model | | |
|---|---|---|---|
| | *Model* | *Feature Set (Scenario)* | *Observed MSE* |
| ROA | Linear Regression | fin-st vars, all vars | 0.00127 |
| ROE | Linear Regression | fin-st vars, all vars | 0.01742 |
| Net Margin | Linear Regression | all vars | 0.05679 |
| Op Margin | Linear Regression | all vars | 0.04816 |
| Cash Ratio | Linear Regression | all vars | 0.27908 |
| Op Cash Generation | SVR | new vars | 6.9191 |

Linear regression models consistently perform better in predicting '*next-quarter financial performance*' across all 6 predictive tasks than all other regressor models considered. Only in OCG prediction, SVR performs better, with LinReg as a close second. Best predictors by task are summarized in **Table 2**. Linear Regression models also show lowest variance in 10-fold training and evaluation (**Fig. A5**).

The introduction of the new 9 financial ratios based on company's data from Statement of Cash Flows and macro-economic data yields an additional uplift in model performance. **Fig. 4** compares performance of models trained with either financial statement variables only (*fin-st vars scenario*) or with all variables (*all-var scenario*) relative to base scenario, which consists of only variables derived from Balance Sheet and Income Statement. While the performance improvement from inclusion of new variables and macroeconomic indexes is negligible in the ROE and Cash Ratio prediction tasks, the proposed new financial ratios deliver an additional 2% improvement in Net Margin and Operating Margin prediction tasks. The uplift is more significant in OCG prediction task: 11% reduction in MSE.

In addition to 6 regression models, 6 different ensemble models were implemented to assess potential model performance improvements:

- **Boosting methods**: AdaBoost, Gradient Boosting (GradB), Histogram Gradient Boosting (HistGB).
- **Bagging methods**: Random Forrest (RF), Extra Trees (ET)
- **Voting method**: a voting regressor consisting of 3 models: a linear regressor, a random forest regressor and a K Neighbours regressor

**Table 3** summarizes results of ensemble models vis-à-vis liner regressor. AdaBoost performs very poorly. Gradient Boosting and Histogram Gradient Boosting perform best, outperforming Linear Regressor by 4 – 6% in ROA, ROE, Net Margin and Op. Margin tasks, and by as much as 14.5% in the OCG task. Only in Cash Ratio prediction task the Linear Regressor performs better than the boosting methods.

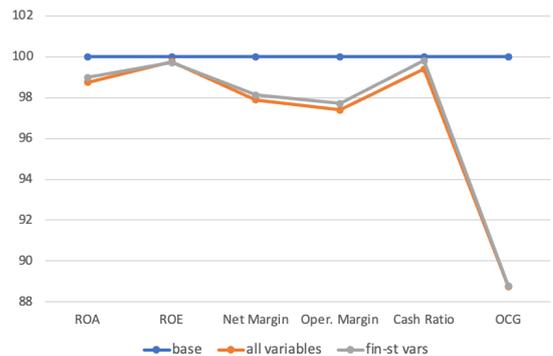

Fig. 4. Linear Regression MSE (indexed, base scenario = 100).

TABLE 3.  MSE FOR ENSEMBLE MODELS – FIN-ST SCENARIO
(33 FINANCIAL RATIOS VARIABLES)

| Predictive Task | Mean Squared Error MSE | | | | | | |
|---|---|---|---|---|---|---|---|
| | Linear Regr. | Ensemble Models | | | | | |
| | | AdaB | GradB | RF | ET | HistGB | Vote |
| ROA | 0.0013 | 0.0020 | 0.0012 | 0.0014 | 0.0014 | 0.0012 | 0.0013 |
| ROE | 0.0174 | 0.0451 | 0.0167 | 0.0187 | 0.0183 | 0.0167 | 0.0172 |
| Net Margin | 0.0569 | 0.0679 | 0.0546 | 0.0604 | 0.0596 | 0.0542 | 0.0582 |
| Op. Margin | 0.0483 | 0.1762 | 0.0456 | 0.0511 | 0.0500 | 0.0454 | 0.0496 |
| Cash Ratio | 0.2802 | 2.0576 | 0.2835 | 0.3230 | 0.3172 | 0.2817 | 0.3282 |
| OCG | 7.2926 | 24.3549 | 6.2345 | 6.6726 | 6.8460 | 6.2149 | 6.4670 |

### B. Neural Networks for Predictive Financial Modelling

Four Feedforward Neural Network (FNN) models were evaluated in the prediction of the same 6 target variables (predictive tasks) previously defined for the ML regression models; and under same 4 scenarios defined by previously described financial ratio feature sets. For all 6 prediction tasks, the FNN Deep and Wide model trained with the financial statement variables (standard balance sheet and income statement ratios, plus new cash-flow and macroeconomic-related financial ratios) performs best. See **Table 4**.

TABLE 4.  FNN – MODEL PERFORMANCE RESULTS

| Predictive Task | Feature Set (Scenario) | Mean Squared Error MSE | | | |
|---|---|---|---|---|---|
| | | NN Base | NN Deep | NN Wide | NN Deep &Wide |
| ROA | all variables | 0.0014 | 0.0013 | 0.0015 | 0.0014 |
| ROA | base | 0.0014 | 0.0013 | 0.0014 | 0.0013 |
| ROA | fin-st vars | 0.0014 | 0.0013 | 0.0014 | 0.0013 |
| ROA | new vars | 0.0013 | 0.0013 | 0.0013 | 0.0013 |
| ROE | all variables | 0.0187 | 0.0183 | 0.0195 | 0.0184 |
| ROE | base | 0.0181 | 0.0176 | 0.0181 | 0.0168 |
| ROE | fin-st vars | 0.0183 | 0.0163 | 0.0180 | 0.0161 |
| ROE | new vars | 0.0201 | 0.0199 | 0.0201 | 0.0195 |
| Net Margin | all variables | 0.0587 | 0.0572 | 0.0585 | 0.0571 |
| Net Margin | base | 0.0592 | 0.0570 | 0.0585 | 0.0556 |
| Net Margin | fin-st vars | 0.0564 | 0.0556 | 0.0562 | 0.0526 |
| Net Margin | new vars | 0.0599 | 0.0599 | 0.0596 | 0.0562 |
| Op. Margin | all variables | 0.0495 | 0.0484 | 0.0501 | 0.0486 |
| Op. Margin | base | 0.0498 | 0.0482 | 0.0497 | 0.0467 |
| Op. Margin | fin-st vars | 0.0477 | 0.0459 | 0.0476 | 0.0438 |
| Op. Margin | new vars | 0.0529 | 0.0522 | 0.0522 | 0.0491 |
| Cash Ratio | all variables | 0.3027 | 0.2932 | 0.3024 | 0.2971 |
| Cash Ratio | base | 0.2978 | 0.2896 | 0.2924 | 0.2771 |
| Cash Ratio | fin-st vars | 0.2902 | 0.2754 | 0.2837 | 0.2746 |
| Cash Ratio | new vars | 0.3666 | 0.3032 | 0.3244 | 0.2830 |
| OCG | all variables | 6.351 | 6.255 | 6.293 | 6.326 |
| OCG | base | 6.424 | 6.308 | 6.396 | 6.093 |
| OCG | fin-st vars | 5.939 | 5.685 | 5.732 | 5.366 |
| OCG | new vars | 6.609 | 6.371 | 6.171 | 5.507 |

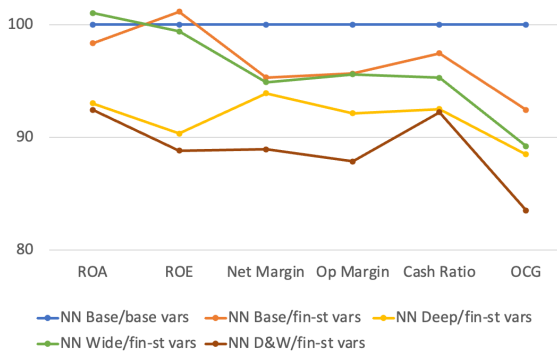

Fig. 5.  Neural Networks MSE (indexed, base scenario = 100)

TABLE 5.  BEST MODELS COMPARISON

| Predictive Task | Mean Squared Error MSE | | | | |
|---|---|---|---|---|---|
| | Base Scen. (24 vars: BS, IS) | Fin. St. Scenario (33 Variables: BS, IS, CF, MFS) | | | |
| | | Regressor | Ensemble | | Neural |
| | Linreg | Linreg | GradB | HistGB | D&W |
| ROA | 0.0013 | 0.0013 | 0.0012 | 0.0012 | 0.0013 |
| ROE | 0.0175 | 0.0174 | 0.0167 | 0.0167 | 0.0161 |
| Net Margin | 0.0580 | 0.0569 | 0.0546 | 0.0542 | 0.0526 |
| Op. Margin | 0.0494 | 0.0483 | 0.0456 | 0.0454 | 0.0438 |
| Cash Ratio | 0.2808 | 0.2802 | 0.2835 | 0.2817 | 0.2746 |
| OCG | 8.2151 | 7.2926 | 6.2345 | 6.2149 | 5.3656 |

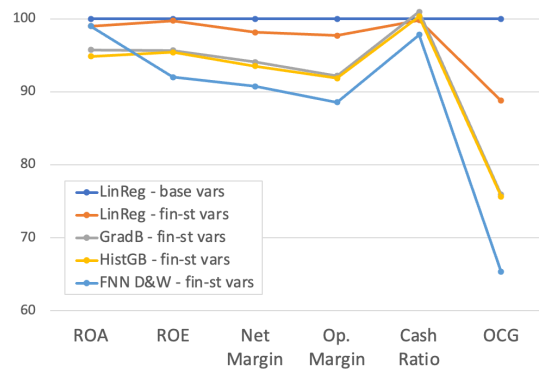

Fig. 6.  Comparative Model Performance (MSE, Indexed to LinReg base = 100)

Compared with a one-hidden layer FNN trained with standard set of BS and IS financial ratios ('*base vars*' scenario), a deeper and wider FNN trained with all financial statement ratios (inclusive of new proposed 9 CF and MRFS ratios) yields 7 – 17% lower error rates across all six predictive tasks (**Fig. 5**). The Deep and Wide FNN model outperformed best ML regressor (LinReg) and Ensemble (GradGB, HistGB) models across 5 of the 6 target variable tasks (**Table 5**). Only in ROA prediction, HistGB performed slightly better. Compared with a supervised learning Linear Regression model trained with standard financial ratios of BS and IS, the deep and wide FNN reduces error (MSE) by 8-11% on ROE, Net Margin and Op Margin predictive tasks, and as much as 35% in the Operating Cash Generation (OCG) predictive task (**Fig. 6**).

### C. Bayesian Network Probabilistic Financial Performance Modelling

The nature of the Bayesian Network model requires a different approach to model performance measurement and evaluation. In supervised learning models, the data is split for training and evaluation. Using a 10-fold training and validation process, 10% of the data is used for performance evaluation at each fold, and results are averaged across all folds. In the Bayesian case, the data is used to determine the NPTs for each node in the Bayesian Network. Given the large size of the sample dataset (i.e., 46k), it is reasonable to assume that mean and standard deviation values are accurate representations of the whole data.

While the model evaluation methods differ between supervised learning and Bayesian models, the same metric is used to evaluate accuracy of predictions (MSE). Thus, strictly speaking, model comparison is not '*apples-to-apples*'. BN model evaluation was done in Agena.ai software, modelling one company at a time. Options to automate evaluation across a large set of companies can be explored in future research.

BNs require a more expert and detail-oriented approach to model design. Thus, effort was limited to one predictive task: Net Margin. Two evaluation experiments were conducted.

First, 11 companies were selected with observed identical target net margin, i.e. companies with different financial ratios in a given quarter during the period 2011-2023, which produced same net margin in the following quarter. For each one, the observed macroeconomic metrics and financial ratios are fed as cases (observed values) and the probabilities of the unobservable latent *Company Performance* node and of the *Target Net Margin* node are recalculated. The expected value for Net Margin from the Bayesian model is compared with the actual observed value to calculate the error. For example, J.M Smucker Co. had a net margin of 9.5% in Q3-19. **Table 6** summarizes the company's financial ratios in the preceding quarter (Q2-19) and the observed macroeconomic metrics. These values are normalized, transformed (range 0:1) and fed into the model as '*hard evidence*'. **Fig. 7** shows recalculated BN node values. The PDF of the outcome node *Target Net Margin* is recalculated given this evidence to determine its expected value (mean): 0.58. This is the min-max normalized expected Target Net Margin value (**Fig. 8**). The corresponding *unnormalized* expected Net Margin is 13.5%. Thus, the error is 0.04 (vs. observed Net Margin of 9.5%), and the MSE is 0.0016.

TABLE 6.   J.M. SMUCKER CO. Q2 2019 FINANCIAL DATA

| Financial Ratios | | Macroeconomic Indexes | |
|---|---|---|---|
| Ratio | Value | Index | Value |
| Current Ratio | 0.72 | Interest Rate [a] | 1.69% |
| Total Asset Turnover | 0.47 | Yield Curve | 0.15% |
| Return on Assets ROA | 0.0126 | Expected Inflation | 1.44% |
| Return on Equity ROE | 0.0261 | Unemployment Rate | 3.6% |
| Op Cash Generation | 4.59 | GDP Change | 3.3% |

[a.] 10-Year Treasury Note

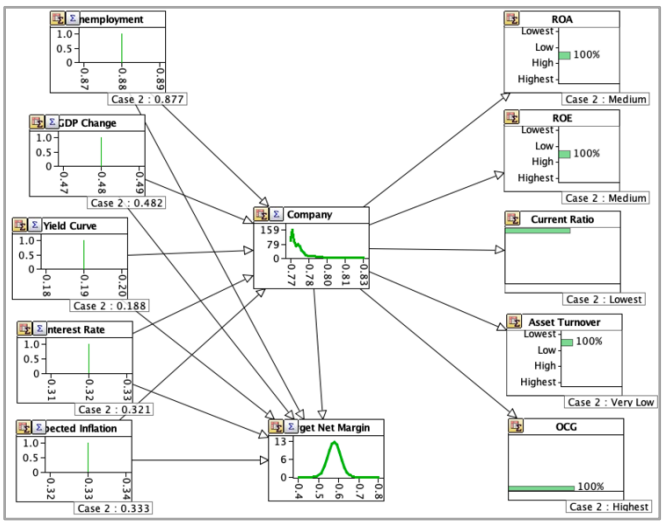

Fig. 7.   J.M. Smucker Co. Bayseian Network – Q2-19.

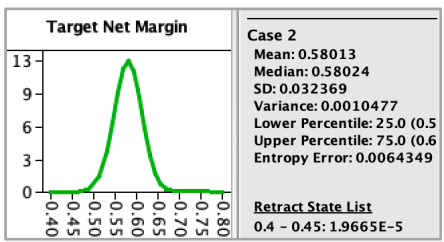

Fig. 8.   J.M. Smucker Co. Expected Net Margin for Next Quarter.

TABLE 7.   MSE – SAMPLE OF 11 COMPANIES WITH OBSERVED NET MARGIN OF 9.5% IN NEXT QUARTER

| Company | BN Model - Target Net Margin ('Next Quarter') | | | | |
|---|---|---|---|---|---|
| | Expected Value (normalized) | Expected Value (un-normalized) | Actual Value (Observed) | Error | Squared Error |
| Alexion Pharmaceuticals Inc. | 0.527 | -11.24% | 9.50% | -0.207 | 0.043 |
| J M Smucker Co. | 0.580 | 13.49% | 9.50% | 0.040 | 0.002 |
| AVX Corp. | 0.546 | -2.16% | 9.50% | -0.117 | 0.014 |
| Warrior Met Coal, Inc. | 0.522 | -13.17% | 9.50% | -0.227 | 0.051 |
| Natus Medical Inc. | 0.551 | 0.14% | 9.50% | -0.094 | 0.009 |
| Rogers Corp. | 0.542 | -4.08% | 9.50% | -0.136 | 0.018 |
| Dorman Products, Inc. | 0.558 | 3.39% | 9.50% | -0.061 | 0.004 |
| Phibro Animal Health Corp. | 0.558 | 3.39% | 9.50% | -0.061 | 0.004 |
| Petmed Express Inc. | 0.543 | -3.86% | 9.50% | -0.134 | 0.018 |
| Turtle Beach Corp. | 0.554 | 1.20% | 9.50% | -0.083 | 0.007 |
| Marine Products Corp. | 0.506 | -20.61% | 9.50% | -0.301 | 0.091 |
| | | | | MSE: | 0.024 |

For the sample of 11 companies with target net margin of 9.5%, the Bayesian model delivers a very high accuracy: average MSE of the sample is 0.024 (**Table 7**), which is lower than 0.053, the average MSE of the FNN neural network model trained with the 46k dataset.

The second experiment widens the sample of companies for which Net Margin is predicted. A subset of 50 companies is randomly sampled from the 46k dataset. Similarly, for each company the Bayesian model is run with the corresponding hard evidence (observed financial ratio values and macroeconomic metrics), to determine the expected value of the outcome node (Target Net Margin). The unnormalized Net Margin values are compared with actual observations. The BN model evidences a drop in performance. Observed average MSE across the 50-companies sample is 0.0585, an 11% increase over FNN Deep and Wide model MSE.

## VII. DISCUSSION

### A. FNNs outperform supervised-learning ML Regressors.

The proposed deep and wide FNN outperformed linear regression models. A LinReg model trained with 33 financial metrics (fin-st scenario comprising of BS, IS, CF and MRFS ratios) has 34 parameters. The FNN model for same scenario has 5,441 parameters. Experimentation shows that with enough training data (i.e., the 46K dataset of templatized financial statements), the FNN model was able to extract more complex relationships between financial metrics, yielding improved prediction accuracy. This is more apparent when predicting financial ratios that are more complex to manage from a CEO standpoint, such as OCG or margin ratios, compared to ROA. In ROA prediction, FNN and LinReg models produced same results.

### B. Under specific circumstances, BN models can outperform FNNs.

The BN Net Margin prediction model performed better than FNN for '*average-performing*' companies (first BN experiment). A practical business forecasting application could first predict financial performance using the FNN model and, upon validating that a company adheres to an '*average-threshold*', calculate probability distributions using the BN model. This approach gets the best from both worlds.

Analysis of predicted vs. observed target Net Margin indicates BN model bias towards underestimating predictions: of the 50 observations (second BN experiment), 31 have negative error and 19 have positive error. The average error is -0.0412. Furthermore, the variance of the predicted values is lower than the variance of actual values (0.0136 vs. 0.0659). This means that errors increase for companies performing farther away from average values and explains why the MSE is low in the sample of 11 companies. This is partially explained by the variances for the macroeconomic nodes and business indicator nodes: once the causal BN is implemented, the underlying variances of the T-Normal distributions calculated by the model are consistently lower (albeit close) than observed data. **Fig. A6** shows the predicted vs. actual Net Margin values for the random sample of 50 companies. Additional experimentation is needed to finetune the parameters of the underlying NPT T-Norm distributions more closely mirroring the variance of the data; and to improve the child node NPT expressions as function of parent nodes.

*C. Domain expertise is very important in predictive financial performance modeling.*

This research shows that feature engineering can drive significant improvement in prediction accuracy. The proposed set of 9 financial metrics (CF and MRFS ratios) demonstrates so. In the case of the ML Regressor models, the impact from new CF and MRFS ratios is almost negligible for simpler target variables (e.g., ROA) and more noticeable for more complex target variables (e.g., OCG). Thus, increasing the dimensionality of the linear regression (vs. base case with 24 variables) does little with simpler financial metrics. Impact from new variables is more significant in case of FNN. The deep and wide FNN has 4,541 parameters in base scenario (24 variables) and 5,441 parameters in the fin-st scenario (33 variables). The later, with more features -and more parameters to learn-, performs better across all 6 predictive tasks.

In case of BN modelling, domain expertise is even more critical as models must be handcrafted, both in terms of network architecture, node (variable) selection and underlying NPT formulae. While the proposed BN model performs well in Net Margin expected value estimation, a different set of macroeconomic nodes and/or business indicator nodes may be more appropriate for other tasks.

*D. Proposed Markovian approach is feasible and makes financial sense – instead of time-series forecasting.*

Time series forecasting is a widely adopted practice in the industry. This assumes that past values are enough to predict future performance. The complexities of running a business have proven such premise to be far from reality. The proposed Markovian approach to predictive business performance modelling is proven to be right. Rather than taking a unidimensional view of a business metric over time, it takes a cross-sectional view into a company's operations. An FNN model that considers 33 financial ratios simultaneously and learns its parameters accordingly is better placed to learn business performance interdependencies. Conversely, an autoregressive moving average (ARMA) model makes linear assumptions of future performance based on past values (*the AR portion*) while modelling the error term as linear combination of error terms at various times in the past (*the MA portion*). It does not attempt to look at interdependencies between multiple variables that may explain variances over time. The downside of FNN is that accurate and standardized data are needed to ensure proper model training. While an ARMA may be constructed with 10 to 15 data points, the FNN model was trained and tested with 46k observations.

## VIII. FUTURE WORK

The present research shows that FNN and BN models have significant potential in predicting a company's financial performance based on its financial statements. Potential is even higher when considering an augmented set of *predictors*, i.e., the proposed ratios based on Statement of Cash Flows and macroeconomic data in addition to the more commonly used ratios based on Balance Sheet and Income Statement.

A progression of this research could focus on stock price prediction. While there is significant research in applicability of neural networks models to predict stock prices, most of it focusses on stock price without deeper attention to the company's foundation. If the financial statements provide an accurate view into a company's *true condition*, theory of efficient markets dictates that the stock price accurately reflects the company's value. However, it is difficult to ascertain that the stock price itself provides enough information to determine if a company is over or undervalued. CEOs are constantly making decisions based on changing business and economic conditions. Predicting future performance based on *just* stock price is like deciding on best move in chess game based solely in the position of the king (*incomplete information*). Predicting future stock price performance based on a more complete set of financial performance ratios tackles the complexity of learning features that look both intrinsically into how a company performs and extrinsically into how the market perceives its performance.

Another venue for proposed future research is the use of Neural Networks to extract features from the Management Discussion and Analysis of Financial Position and Results of Operations (MD&A) and the Notes to the Financial Statements. These are an important part of the overall financial disclosure by a company, providing management with "flexibility to describe the financial matters impacting the registrant" (SEC Financial Reporting Manual). An analysis that focuses only on numerical data may not be as effective. Hybrid Transformer architectures can be explored to encode text data (extract features) and decode it into financial performance metrics (*observed* values during model training, and *predicted* values during model evaluation).

The Markov property assumption can also be challenged. Recurrent Neural Networks (RNN) can be explored as an alternative to proposed FNNs. Use of RNN such as LSTM can be used to capture temporal dynamic behaviours in financial performance during previous quarters (*t-3, t-2, t-1*) in addition to current quarter (*t*) to potentially improve financial performance predictions (*t+1*).

Research on use of BNs can be expanded along two axes: (1) improving the proposed Net Margin prediction model by modifying the node NPT expressions and weights used to estimate child node mean and variance parameters as function of parent nodes, modifying the T-Normal Parameters in order to more closely resemble actual data (parent nodes), or even using Bayesian statistical inference to learn the parameters of each variable distribution (instead of using classical frequentist approach that takes the mean and variance of the sample data as parameter estimates). And (2) designing new models to predict other target variables (ROE, EBIT, etc).

Lastly, additional work can be conducted in expanding set of financial performance ratios used in model training and prediction. While this research conclusively demonstrates the value from the proposed new 9 financial ratios, it cannot be concluded that there is no room for improvement. Examples of new ratios to explore may include company interest expenses, tax payments, Cost of Goods Sold, research expenses, sources / uses of Investing and Financing Cash Flows, etc. This would require revisiting the proposed financial statement templatization and perhaps consider industry-specific templates.

APPENDIX

*A. Tables*

TABLE A1.  FINANCIAL RATIOS FROM BALANCE SHEET AND INCOME STATEMENT DATA

|  | *Ratio* | *Formula* | *Description* | *Observations* |
|---|---|---|---|---|
| **Liquidity Ratios** | Current Ratio | CA / CL | Measures company's liquidity, i.e., ability to cover short term liabilities. | Industry rule of thumb: Current Ratio > 1 means company is liquid. However, benchmark varies by industry. |
|  | Quick Ratio | (CA-Inv) / CL | A modified version of the Current Ratio - it excludes inventories from the calculation, as inventories are more difficult to liquidate. | Reduction in inventories can improve this metric. However, should inventory levels fall too short may negatively impact sales. |
|  | Cash Ratio | cash / CL | Measures ability to cover current liabilities strictly against current cash position. | It is the most stringent liquidity KPI, as it only takes into consideration cash to measure liquidity (ability to cover short term liabilities). However, too much cash is not optimal as it may become an unproductive asset. |
| **Asset Utilization Ratios** | Total Asset Turnover | sales / assets | Measures company's efficiency in utilizing scarce resources (assets). Specifically, revenues being generated related to the assets employed. | It provides insights into how much total asset is needed for the company to generate revenues. |
|  | Fixed Asset Turnover | sales / fixed assets | Measures revenues being generated related to the fixed assets (long term). | A tighter version of the previous KPI, as it focusses on fixed asset utilization. |
|  | Tangible Asset Turnover | sales / (fixed assets - intangibles) | A modified version of the previous KPI, but it excludes intangible assets, which valuations may be subjective. | Decreases on asset turnover ratios indicate company using more fixed assets to generate same amount of revenues. Deducing intangibles provides a more accurate perspective on company's efficiency. |
|  | PPE Turnover | sales / PPE | Most stringent version of asset turnover ratios, as it only considers productive PPE assets. |  |
|  | Days Sales Outstanding (DSO) | AR / (sales / 365) | Indicates how long it takes a company to collect the funds from accrued sales. | Measures efficiency of the collections function. In quarterly data analysis, use sales / (365*/4). |
|  | Accounts Receivable Turnover | sales / AR | Inverse of Days Sales Outstanding. It measures how many times a company is able to collect funds from customers. | A company may be able to increase sales without getting paid in timely manner. This KPI provides visibility into efficiency in ensuring revenues are timely collected. |
|  | Average Days Payable (ADP) | AP / (sales / 365) | Indicates how long (in average) it takes a company to pay to its suppliers. | A company that pays too quickly shows suboptimal utilization of its working capital. Conversely, a company that pays late risks relationships with suppliers. Prompt payment could be linked to incremental discounts from suppliers, and thus increased margins. However, this KPI cannot independently assess impact on profitability. |
|  | Accounts Payable Turnover | sales / AP | Measures how many times a company paid out its average accounts payable amount on an annual (or quarterly) basis. It is the inverse of Average Days Payable. | Inverse of Average Days Payable. |
|  | Inventory Turnover | sales / inventory | Measures how many times a company is able to cycle through the inventory stock. | Indicates how good a company is at keeping inventories low and replenishing them if and when needed. |
|  | Average Days in Inventory (ADI) | inventory / (sales / 365) | Measures how long a company takes to sell or move its inventory. It is the inverse of Inventory Turnover. | Inverse of Inventory Turnover . |
|  | Working Capital Turnover | sales / WC | Indicates how well a company exploits working capital opportunities. | Higher ratios indicate increased ability to generate sales with lower working capital. |
| | Cash Cycle | ADI + DSO - ADP | Indicates, in days, the length of the average business cycle to produce cash, discounting the time taken to pay suppliers. | |
| **Profitability Ratios** | ROA Return On Assets | NI / assets | Measures profitability against total assets utilized by the company. | Higher ROA indicates ability to make more profits with same (or less) assets. It differs from Asset Turnover as it focuses on the bottom line (net income). |
| | ROE Return On Equity | NI / equity | Measures profitability against equity invested. | A preferred metric by Investors. Besides its absolute value, investors value ROE trending upwards and thus, it can be argued that future ROE growth is a key company management objective. However, it cannot be analysed in isolation: a company may artificially increase ROE by increasing debt to finance operations instead of equity, potentially putting company at unsustainable leverage levels. |
| | Net Profit Margin | NI / sales | Compares company's net income with its revenues: for each dollar of revenues, it measures how much is available to investors as profit. | High margins indicate efficient production and supply chains, low overhead and low financing costs. One minus net profit margin determines how much money is spent on operations and financing. |
| | Gross Profit Margin | (revenues - COGS) / revenues | Measures how much funds are available after deducting direct costs associated with the production of goods or delivery of services. | High margins indicate efficient production and supply chains. |
| | Operating Profit Margin | EBIT / sales | Measures how much funds are available after deducting direct costs associated with the production of goods or delivery of services and indirect (overhead) expenses. | High margins indicate efficient production and supply chains, as well as low overhead. This ratio helps excluding management actions that can increase profit margins through non-core operations such as sale of investment securities. |
| | Basic Earning Power | EBIT / assets | Similar to ROA but excludes non-core income sources (or uses) from the calculation, to provide a profitability measurement more aligned with core operations. | |
| **Debt Ratios (Financial Leverage)** | Debt Ratio | liabilities / assets | Measures what percent of the total assets is funded with debt. | Companies with high Debt Ratio are highly leveraged and thus, have higher risk exposure (to increases in interest rates). |
| | LT Debt Ratio | LT liabilities / assets | Measures what percent of the total assets is funded with long term debt. | Companies with relatively larger Debt Ratio compared to LT Debt Ratio mean that they have comparatively larger short-term liabilities. |
| | Debt to Equity Ratio | LT Debt / common equity | Indicates how much debt is used compared to equity. | A 1:1 ratio means that the company uses lenders and investors in equal proportion. |

| | |
|---|---|
| AP | Accounts Payable |
| AR | Accounts Receivable |
| CA | Current Assets |
| CL | Current Liabilities |
| COGS | Cost of Goods Sold |
| EBIT | Earnings Before Interest and Taxes |
| Inv | Inventory |
| LT | Long-term |
| NI | Net Income |
| PPE | Property, Plant and Equipment |
| WC | Working Capital (WC = CA - CL) |

TABLE A2.   SEC FILINGS BY INDUSTRY (JANUARY 2011 – APRIL 2023)

| *Industry* | *Companies* | *SEC filings* | *10Q* | *10K* | *10Q+10K* |
|---|---|---|---|---|---|
| Industrial Applications and Services | 1310 | 54275 | 23655 | 7883 | 31538 |
| Energy & Transportation | 2000 | 64989 | 28476 | 9377 | 37853 |
| Life Sciences | 1284 | 56103 | 20341 | 6917 | 27258 |
| Manufacturing | 1935 | 75862 | 33359 | 11007 | 44366 |
| Real Estate & Construction | 2811 | 65588 | 29658 | 9670 | 39328 |
| Technology | 1861 | 62204 | 26859 | 8871 | 35730 |
| Trade & Services | 2454 | 81705 | 37490 | 12286 | 49776 |
| **Grand Total** | **13655** | **460726** | **199838** | **66011** | **265849** |

TABLE A3. NEW FINANCIAL RATIOS PROPOSED IN CURRENT RESEARCH

New set of metrics (ratios) based on Statement of Cash Flows and Macroeconomic indexes, which supplement standard and commonly used financial ratios based on Balance Sheet and Income Statement data.

| Sources and Uses of Cash Ratios | |
|---|---|
| **Net Cash Generation** *NCG* | Measures how much of the outstanding cash at the end of the period was generated from either operations, financing or investing activities during the period. Values greater than 1 indicate that cash on balance has increased during the period, although the metric does not differentiate source of cash (operations, investing or financing). |
| **Operating Cash Generation** *OCG* | Measures how much of the outstanding cash at the end of the period was generated from operations during the period. OCG values greater than +1 measure how many times the company produces cash on Balance Sheet during the period. OCG values between 0 and 1 indicate that company produces net operating cashflows, but other cash in balance sheet is coming from investing and / or financing activities. Negative OCG values indicate that operations are net cash negative, and cash on Balance Sheet comes strictly from investing and / or financing activities, representing a riskier business performance. OCG metric is best when considered in tandem with other financial ratios such as Current Ratio since a low cash on balance yields a higher OCG value. |
| **Current Liabilities Cash Coverage** *CLCC* | Measures how many times current liabilities are covered by operating cashflows produced during the period. Positive values greater than 1 indicated good ability to produce cash to cover short term liabilities. Positive values in range 0 to 1 indicate that while the company produces positive operating cash flows, these are not enough to cover short term liabilities and thus, other sources of cash (investing and / or financing) would be required. Negative CLCC values indicate that Current Liabilities would need to be covered by financing or investing sources. |
| **Operating Cash to Sales Ratio** *OCS* | Measures the ratio between cash flow produced from operations and sales during the same period. In other words, it measures how many dollars of operating cash are generated for each dollar of sales. |
| **Quality of Payment Terms** *QPT* | Measures the quality of the collection function. f1 is a non-linear transformation of DSO (Days Sales Outstanding) to better assess quality of payment terms. Companies that are paid within 30 days or less are scored at 1. Companies with DSO greater than 90 days are scored at -1. Companies that are paid in average within 30 to 90 days get a score between +1 and -1, with QPT = 0 at 60 days. QPT equally values companies that collect within 30 days at +1, or later than 90 days at -1, irrespective of the actual number of days. A company that collects at 100 days is deemed equally sub performing than a company that collects at 120 days. |
| **Quality of Op to Fin Funds Use Ratio** *QOFFUR* | Score in the range -1 to +1. It measures the quality of the cash flow from operations relative to cash flow from financing activities. Both positive and negative values of operating and financing cashflows are acceptable or questionable under different circumstances. Thus, QOFFUR takes a more nuanced approach to assess the quality. At one end, a company that produces operating cashflows (CFop > 0) and uses part of it to pay lenders back (CFfin <0) gets a score of 1. A company that has net negative operating cashflows and net negative financing cashflows is a company that continues to borrow without being able to produce positive cashflows. It gets a score of -1. QOFFUR values between +1 and -1 address different permutations of CFop and CFfin polarities and values. |
| **Macroeconomic-Related Financial Statement Ratios (MRFS)** | |
| **Liabilities to Yield Curve Alignment** *LYCA* | Measures the alignment between the debt structure of the company and the macroeconomic interest rates determined by the Yield Curve. Positive values indicate a better alignment of the company's short-term vs. long-term debt with market ST/LT rates. Negative values indicate misalignment. For example, a company with 80% LT debt and 20% ST debt when Yield Curve = -1% (meaning ST rates are higher than LT rates by 1%) would yield a LYCA value of +0.006. The company debt structure of favours LT debt, which has lower rates. Thus, company's LYCA value is positive, indicating good alignment. |
| **Inflation-Adjusted Inventory Carry-on Cost** *IAICOC* | Measures the economic cost of carrying inventories given expected inflation rates and company's inventory turnover. In inflationary situations, the company is penalized if carrying larger inventories relative to assets. The IAICOC value is a function of the company's inventory turnover (average days in inventory), its inventory levels relative to assets, and the inflation rate. |
| **ROA to Bond Rate Ratio** *ROA2bond* | Measures company's Return on Assets relative to the prevalent average cost of borrowing determined by Moody's BAA bond rate. It is intended to measure if a company yields better returns than cost of borrowing. |

## B. Figures

| Balance Sheet | Statemet of Cash Flows |
|---|---|
| Cash | NetIncome_CF |
| Inventory | BS_adjust |
| Acc_Receivable | IS_adjust |
| AssetsCurrent_Other | CF_Op_Total |
| AssetsCurrentTotal | |
| AssetsCurrentTotal_CALC | CF_Op_Total_CALC |
| chk01 | chk08 |
| Property | CF_Inv |
| Intangibles | CF_Inv_Total |
| AssetsNonCurrent_Other | |
| AssetsTotal | chk09 |
| | CF_Fin |
| AssetsTotal_CALC | CF_Fin_Total |
| chk02 | |
| AccPayable | chk10 |
| NotesIntConvDebt | CF_Other |
| OtherAccruedLiab | CF_Net_Total |
| LiabilitiesCurrent_Other | |
| LiabilitiesCurrentTotal | CF_Net_Total_CALC |
| | chk11 |
| LiabilitiesCurrentTotal_CALC | |
| chk03 | **Income Statement** |
| LiabilitiesNonCurrentOther | Net_Revenues |
| LiabilitiesCurrNonCurrOther | Op_Income |
| LiabilitiesTotal | Net_Income |
| LiabilitiesTotal_CALC | chk12 |
| chk04 | |
| PaidInCapital | |
| RetEarnings | |
| TreasuryStock | |
| Equity_Other | |
| EquityTotal | |
| EquityTotal_CALC | |
| chk05 | |
| EquityOther_Minority | |
| EquityMinorityTotal | |
| EquityMinorityTotal_CALC | |
| chk06 | |
| LiabilitiesAndEquityTotal | |
| LiabilitiesAndEquityTotal_CALC | |
| chk07 | |

Fig. A1. Standardized Financial Statement Template. Tags in grey background correspond to accounts extracted from the 10Q, 10K filings. Tags in blue named *[field_name]_CALC* are used for fields that are calculated using the corresponding subaccounts. Tags in blue named *chk0x* are checkpoints used to compare extracted parent account data with values calculated from corresponding child subaccount data, to verify the accuracy of the data extraction and classification code.

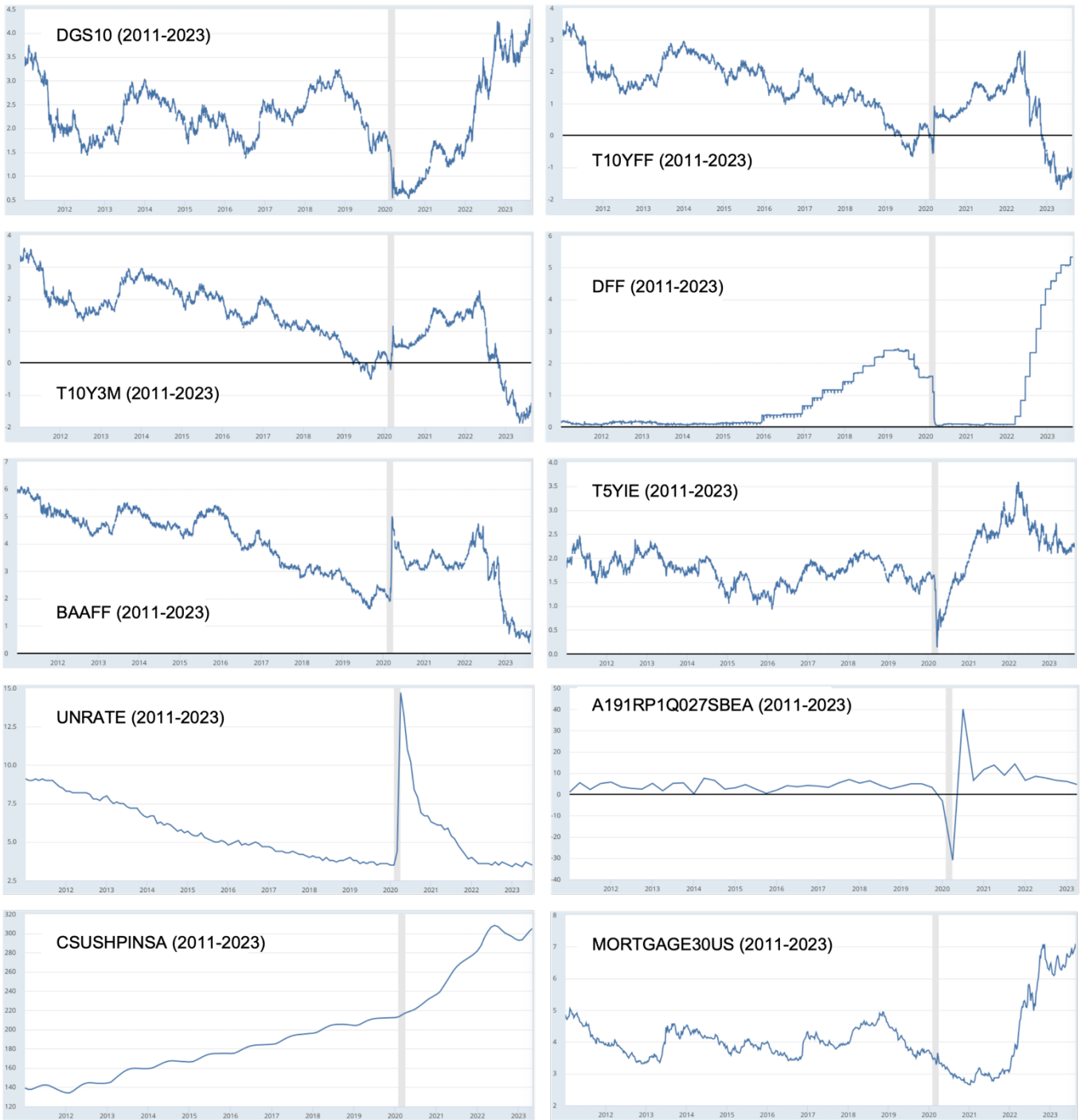

Fig. A2. Macroeconomic Indexes 2011-2023. DGS10: Market Yield on U.S. Treasury Securities at 10-Year Constant Maturity, Quoted on an Investment Basis, Percent, Not Seasonally Adjusted. T10YFF: 10-Year Treasury Constant Maturity Minus Federal Funds Rate, Percent, Not Seasonally Adjusted T10Y3M: 10-Year Treasury Constant Maturity Minus 3-Month Treasury Constant Maturity, Percent, Not Seasonally Adjusted. DFF: Federal Funds Effective Rate, Percent, Not Seasonally Adjusted. BAAFF: Moody's Seasoned Baa Corporate Bond Minus Federal Funds Rate, Percent, Not Seasonally Adjusted. T5YIE: 5-Year Breakeven Inflation Rate, Percent, Not Seasonally Adjusted. UNRATE: Unemployment Rate, Percent, Seasonally Adjusted. A191RP1Q027SBEA: Gross Domestic Product, Percent Change from Preceding Period, Seasonally Adjusted Annual Rate. CSUSHPINSA: S&P/Case-Shiller U.S. National Home Price Index, Index Jan 2000=100, Not Seasonally Adjusted. MORTGAGE30US: 30-Year Fixed Rate Mortgage Average in the United States, Percent, Not Seasonally Adjusted. Source: U.S. Bureau of Economic Analysis, retrieved from FRED, Federal Reserve Bank of St. Louis; https://fred.stlouisfed.org, 19 August 2023.

Fig. A3. Correlation Matrix of set of financial ratios (variables) used in FNN and ML Regressor models. Metrics in light grey background correspond to 24 standard financial ratios based on Balance Sheet and Income Statement. Metrics in blue background correspond to 6 new proposed metrics based on Statement of Cash Flows. Metrics in green background correspond to 3 new proposed Macroeconomic Related Financial Statement ratios (MRFS), which relate data from financial statements to macroeconomic indexes. Data in orange background correspond to 10 macroeconomic indexes. Data in purple background correspond to the six target variables used in predictive tasks. Numerical values with navy blue background show correlation values closer to +1. Numerical values with red background show correlation values closer to -1.

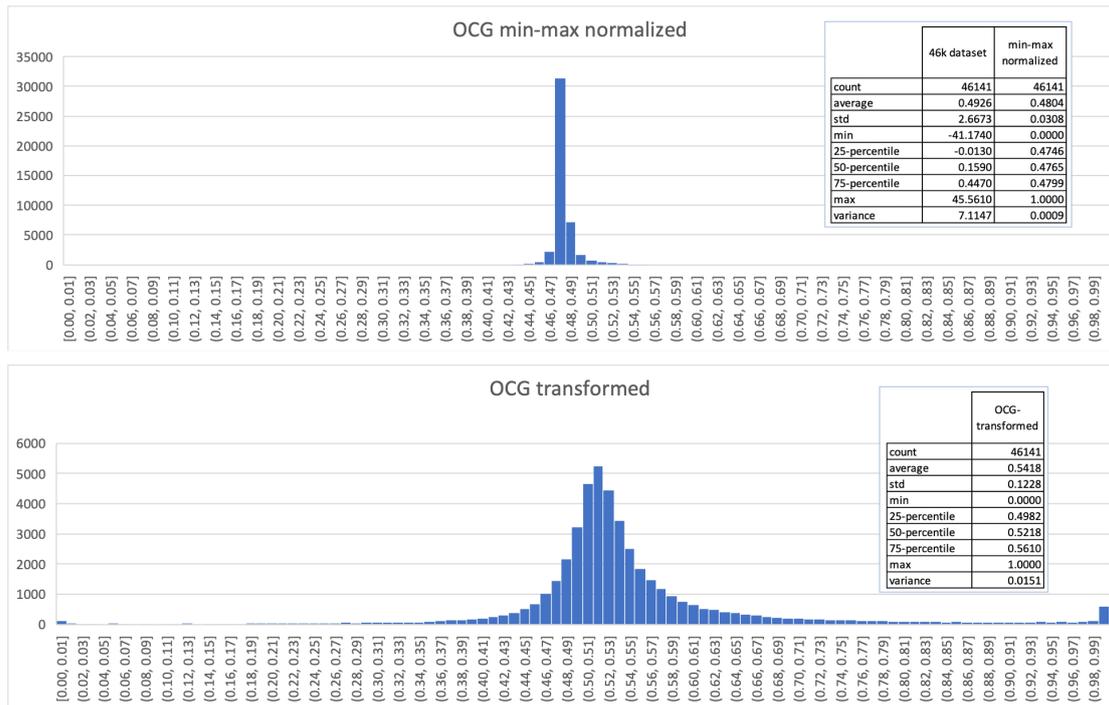

Fig. A4. Operating Cash Generation Ratio (OCG). Top chart: Histogram of OCG ratios for 46k records. Bottom chart: histogram of OCG': a transformation of OCG ratios to augment value discerning in production of the ranked node NPT. The values in the 46K training dataset are in the range -41 to +46, which are then normalized to fit 0:1 range. Large absolute values are characteristic of business transient states and not sustainable over time. Thus, the shape of the distribution histogram is very narrow with very long tails (i.e., very low variance). Use of ranked nodes on normalized data without proposed transformation would place most of the companies in the middle bucket, not allowing for proper differentiation across companies.

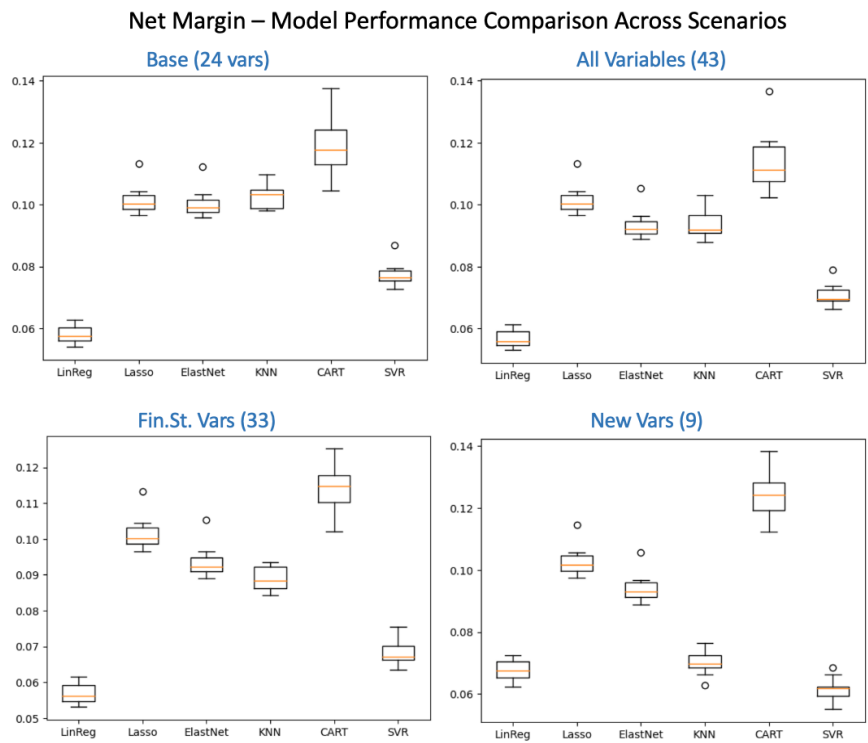

Fig. A5. Box diagrams comparing model performance (MSE) on 10-fold evaluation of Net Margin prediction for the proposed 6 regression models across 4 scenarios. LinReg models consistently deliver lowest MSE and lowest variance across 10 folds, as illustrated by the shorter box without outlier observations. Only SVR performs marginally better in the New-Vars scenario (models trained only with new proposed CF and MRFS financial ratios).

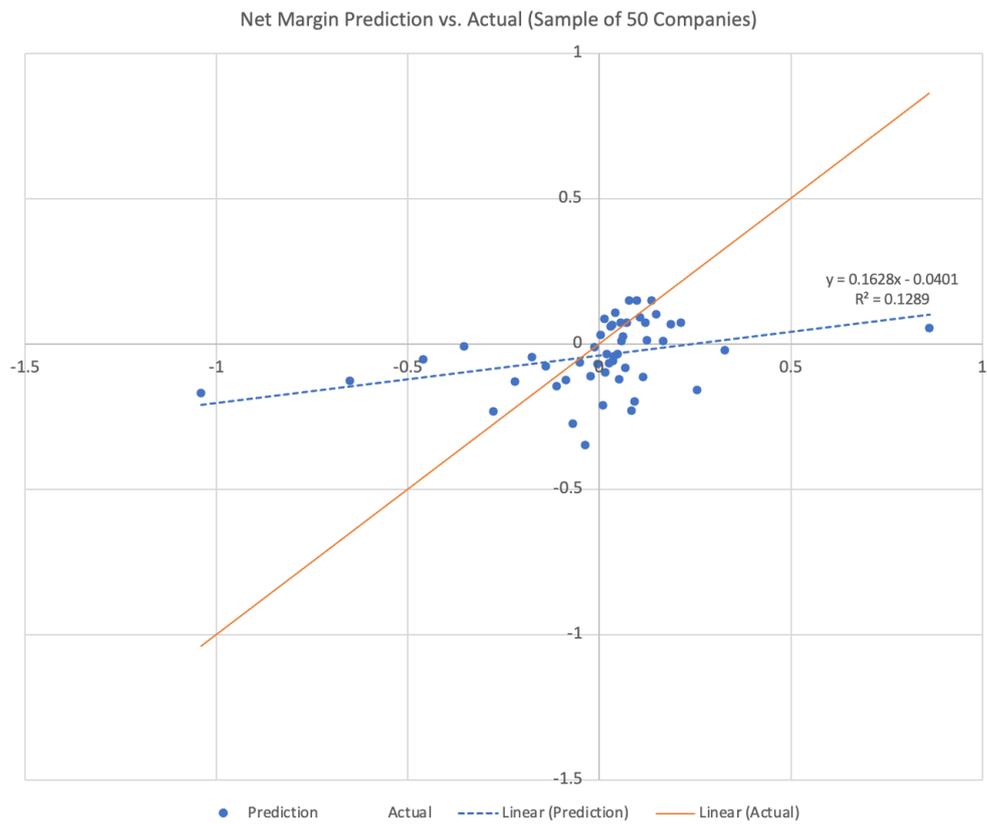

Fig. A6. Net Margin: Prediction (Y axis) vs. Actual (X axis). Predicted value corresponds to expected value from Bayesian Network model, calculated upon entering *hard evidence* corresponding to financial ratios calculated for each company in the sample. Random sample of 50 companies.